\documentclass[aps,pra,twocolumn,groupedaddress]{revtex4}

\usepackage{graphicx}
\usepackage{dcolumn}
\usepackage{bm}

\begin{document}


\title{Dynamics of the vortex-particle complexes bound to the free surface of superfluid helium}



\author{P. Moroshkin$^{1,2}$}
\email[]{petr.moroshkin@oist.jp}
\author{P. Leiderer$^{3}$}
\author{K. Kono$^{1,4}$}
\author{S. Inui$^{5}$}
\author{M. Tsubota$^{5}$}

\affiliation{$^{1}$RIKEN Center for Emergent Matter Science, 2-1 Hirosawa, Wako, 351-0198 Saitama, Japan}

\affiliation{$^{2}$Okinawa Institute of Science and Technology, 1919-1 Tancha, Onna-son, 904-0495 Okinawa, Japan}

\affiliation{$^{3}$Department of Physics, University of Konstanz, Universit\"{a}tstrasse 10, 78464 Konstanz, Germany}

\affiliation{$^{4}$Department of Electrophysics, National Chiao Tung University, Hsinchu 300, Taiwan}

\affiliation{$^{5}$Department of Physics, Osaka City University, 3-3-138 Sugimoto, 558-8585 Osaka, Japan}

\date{\today}

\begin{abstract}
We present an experimental and theoretical study of the 2D dynamics of electrically charged nanoparticles trapped under a free surface of superfluid helium in a static vertical electric field. We focus on the dynamics of particles driven by the interaction with quantized vortices terminating at the free surface. We identify two types of particle trajectories and the associated vortex structures: vertical linear vortices pinned at the bottom of the container and half-ring vortices travelling along the free surface of the liquid.
\end{abstract}

\pacs{xxxx}

\maketitle


Quantized vortices are topological defects that exist in superconductors and in superfluids, such as superfluid helium, atomic Bose-Einstein condensates and exciton-polariton condensates.
Vortices in superfluids can be created by microscopic impurity particles \cite{MuirheadPTRSLA1984} and by various mesoscopic objects \cite{BradleyPRL2000,BlazkowaPRB2009,KwonPRA2015} moving faster than a certain critical speed.
Suspended microparticles interact with vortices and become bound to them \cite{KivotidesPRB2007,KivotidesPRB2008,ShuklaPRA2018}.
Free motion of these particle-vortex complexes has been observed in bulk superfluid $^{4}$He \cite{BewleyN2006,GordonJETPL2007,BewleyJLTP2009,FondaRSI2016}.
Related effects have also been studied in superfluid He nanodroplets containing impurity atoms and nanoparticles \cite{GomezS2014}.

Dynamics of a free surface is another important topic in the research on quantum fluids.
The free surfaces of superfluid $^{4}$He and $^{3}$He have been investigated using free electrons and He$^{+}$ ions as probes.
Both types of charged particles can be localized at the free surface and driven parallel to it by the external electric field.
This approach has led to the observations of anomalous Hall effect of topological origin in superfluid $^{3}$He-A \cite{IkegamiS2013,ShevtsovPRB2016}, of Majorana surface states in superfluid $^{3}$He-B \cite{IkegamiJPSJ2013,TsutsumiPRL2017}.

The research on quantum hydrodynamics has a long history \cite{TsubotaPR2013}, however most efforts have been devoted to bulk.
Until recently, little attention has been paid to the phenomena involving quantized vortices near solid boundaries or surfaces \cite{YuiPRL2018}, in particular a free surface.
Here we present a new experimental and theoretical study of the motion of particle-vortex complexes in superfluid He.
Our experiments visualize the motion of the tip of a vortex terminating at a free surface.
The vortex is bound to an electrically charged particle which is trapped under the free surface due to the applied electric field and the surface tension.
Our observations combined with the numerical modeling allow us to identify two types of vortices that form particle-vortex complexes under these conditions: a linear vortex stretched in a vertical direction and a half-ring vortex with both ends terminating at the free surface.

\begin{figure} [tbp]
	\includegraphics [width=\columnwidth] {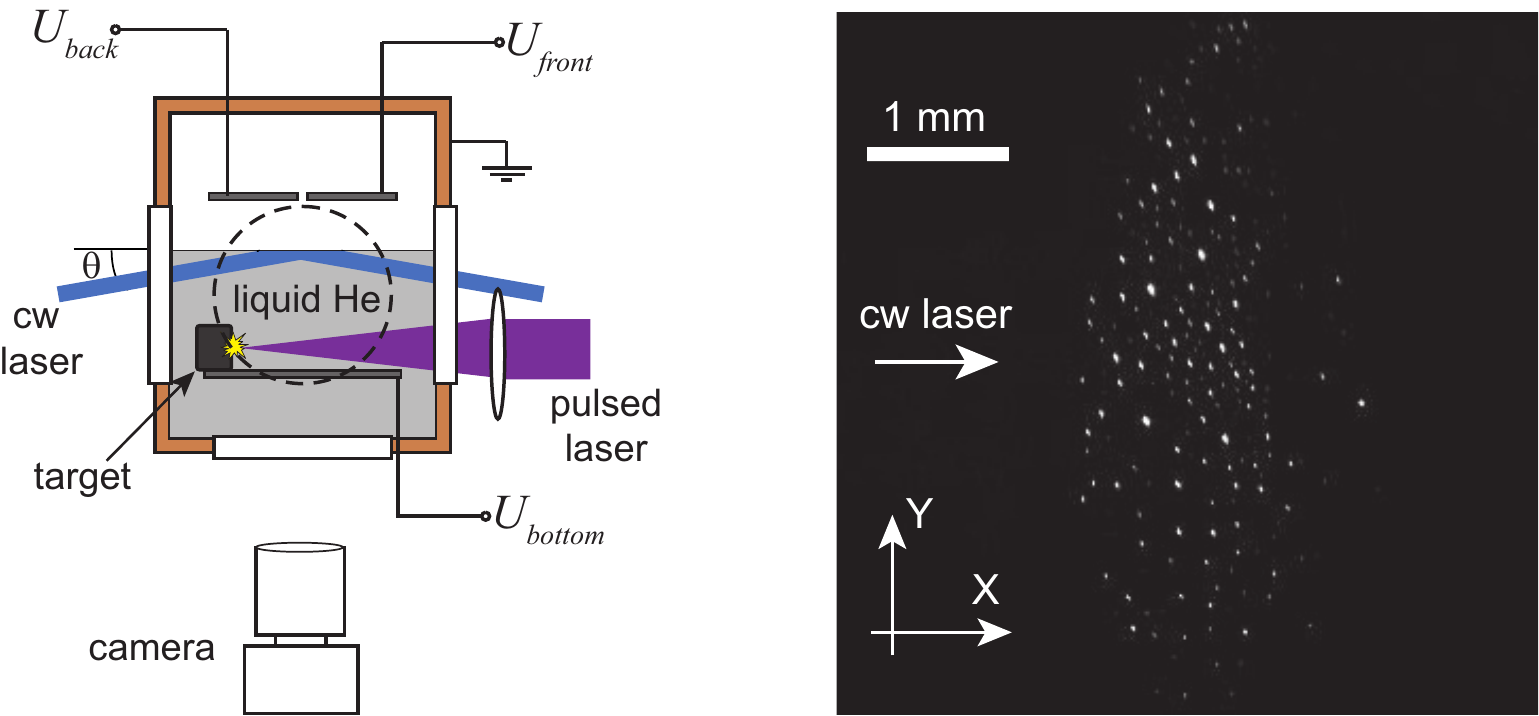}
	\caption{Left - experimental sample cell; right - typical image of trapped particles (bottom view).}
	\label{fig:Setup}
\end{figure}

Our experiments are performed in a helium-bath cryostat with optical access via four side-windows and a window in the bottom.
The temperature is adjusted in the range of $T$ = 1.35 - 2.17 K by pumping on the liquid He in the bath.
The sample cell is immersed in the He bath and is filled up to a certain height by condensing He gas from a high pressure gas cylinder.
A system of three horizontal flat electrodes shown in Fig. \ref{fig:Setup}(a) is used to create a static electric field, with a predominantly vertical orientation.
The bottom electrode is transparent and thus allows us to monitor the interior of the cell and in particular the liquid He surface via the window at the bottom.

The tracer particles are produced inside the cell and are trapped at the free surface by the technique developed in our recent studies \cite{MoroshkinPCCP2016,MoroshkinPF2017,MoroshkinPRE2017}.
A frequency-tripled pulsed diode-pumped solid-state (DPSS) laser ($\lambda$ = 355 nm) with a pulse energy of 70 $\mu$J is used to ablate a metal (Cu, Ba, Dy) target submerged in liquid He.
A single laser pulse is sufficient to produce several tens or hundreds of metallic micro- and nanoparticles which become electrically charged either in the ablation plume, or by touching the bottom electrode.
The charged particles then rise up and become trapped under the free surface of the liquid He.
The trapping potential is provided by the electric field that pulls the particles upwards and by the surface tension of liquid He which does not allow them to cross the liquid-gas interface.

In order to visualize the injected particles, we use scattered light from the beam of a cw frequency-doubled diode laser with a wavelength $\lambda$ = 480 nm.
The laser beam is expanded in a horizontal direction and illuminates the particles via a side window.
The motion of particles along the surface is recorded by a digital video camera, installed under the bottom window of the cryostat and operated at a frame rate of 100--5300 fps.
The angle $\theta$ between the cw laser beam and the free surface of the liquid is adjusted in the range of $\pm10^{\circ}$ in order to get rid of dark zones appearing due to surface waves which are excited by slight vibrations of the experimental set-up.

The size of each individual particle is not known.
The scanning electron microscope (SEM) study \cite{MoroshkinPCCP2016} of the deposites collected after the experiment has revealed two types of particles: spheres with diameters in the range of 20--500 nm and nanowires with a uniform diameter of $\approx$10 nm and a broad distribution of lengths. 
Our estimates \cite{MoroshkinPCCP2016} confirm that Ba particles of such sizes, under the conditions of our experiments scatter a sufficient amount of light to be detected by the camera.

At constant voltages applied at each electrode and with the ablation laser switched off, the particles form a quasi-static array, as shown in Fig. \ref{fig:Setup}(b).
The system of electrodes shown in Fig. \ref{fig:Setup}(a) allows us to apply an alternating horizontal electric field and thus drive the motion of particles along the surface.
These results will be presented in a forthcoming publication, here we discuss only the free motion of the particles under a constant electric field.

\begin{figure} [tbp]
	\includegraphics [width=\columnwidth] {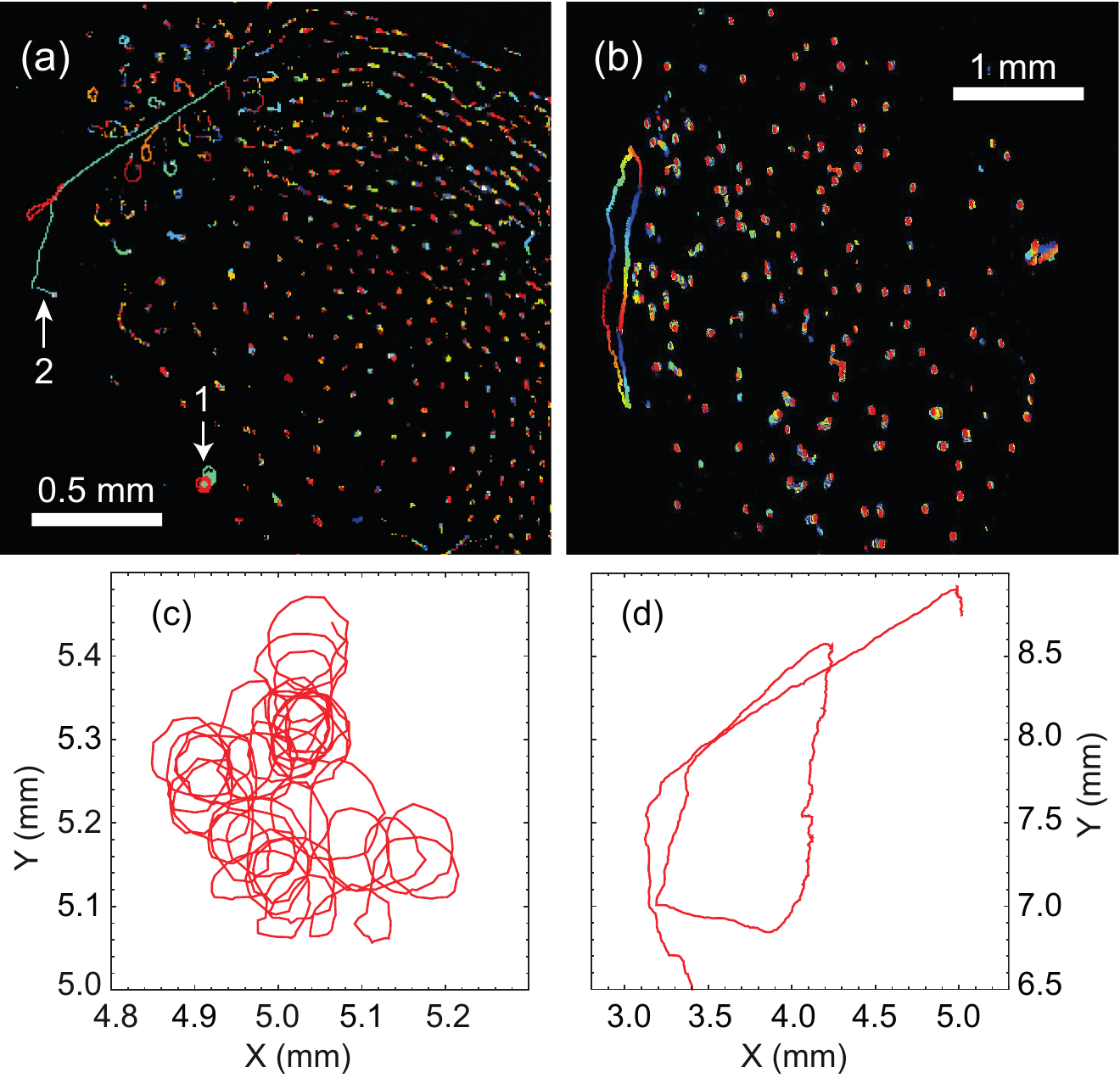}
	\caption{(a), (b) Maps of particle trajectories. (c), (d) Traced trajectories of two anomalous particles from (a). Experimental conditions in (a), (c), (d): $T$ = 1.35 K, $U_{front}$ = +100 V, $U_{bottom}$ = -110 V, cw laser power 42 mW; in (b): $T$ = 2.1 K, $U_{front}$ = -116 V, $U_{bottom}$ = 0, cw laser power 35 mW.}
	\label{fig:MapsTraj}
\end{figure}

In equilibrium, the particles occupy the positions at the free surface where the component of the local electric field along the surface (\textit{i.e.} horizontal) is equal to zero.
Surface waves tilt the free surface locally, thus producing at the particle location a non-zero component of the electric field parallel to the surface.
This leads to collective particle oscillations in the $XY$ plane that is nicely resolved by our technique.
All particles within the same array oscillate in phase, with a coherence time of several seconds and with an amplitude of 50 $\mu$m, or less.
Under quiet conditions, the oscillations become barely visible.
In Figs. \ref{fig:MapsTraj}(a) and (b) we show the maps of particle trajectories from two different video recordings, obtained by tracing the motion of all particles in each recording with the help of $Diatrack$ \cite{VallottonT2017} software.
All regular particles are represented as well separated blobs, their size reflecting the amplitudes of the particle oscillations along $X$ and $Y$ directions.

\begin{figure} [tbp]
	\includegraphics [width=\columnwidth] {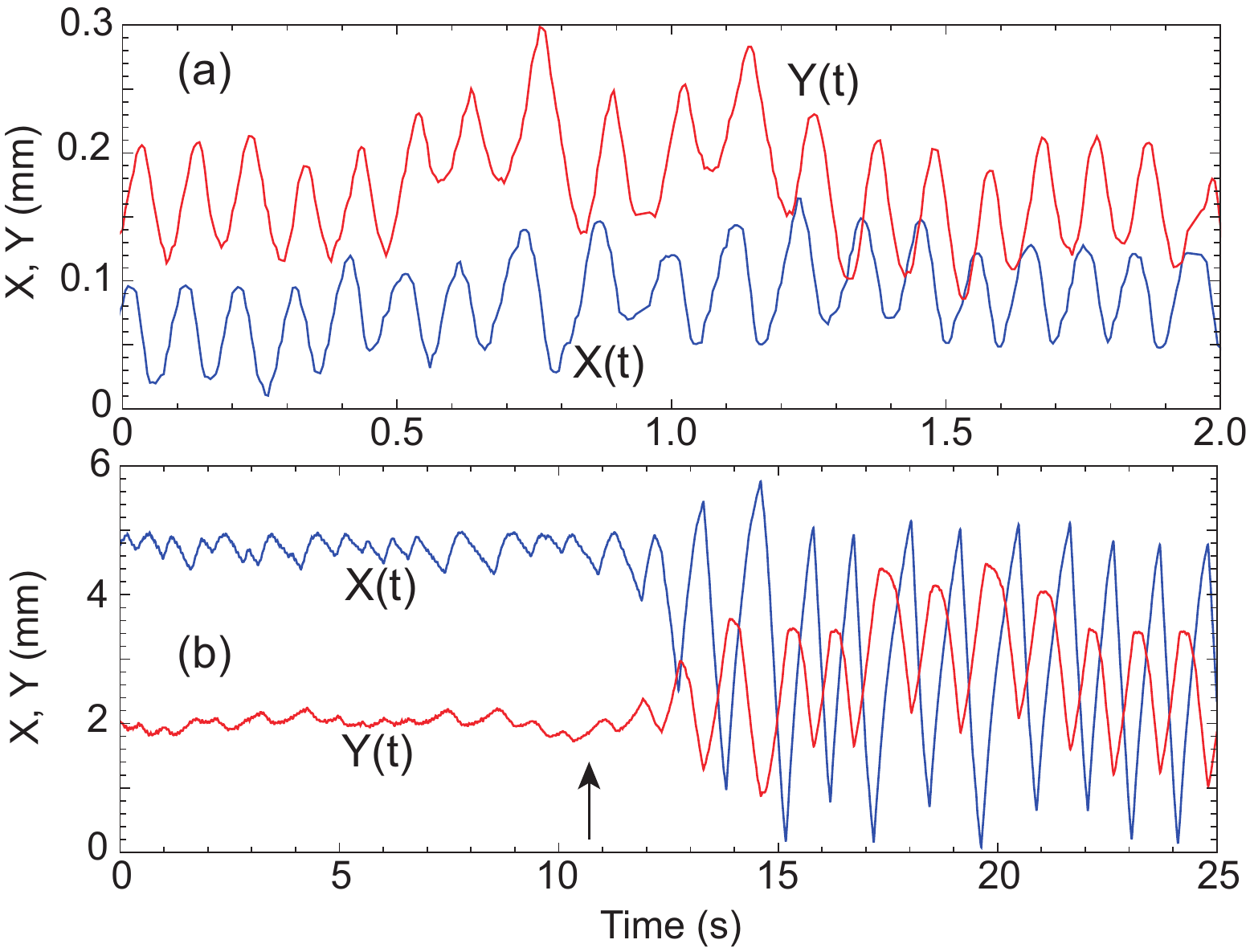}
	\caption{$X(t)$ and $Y(t)$ coordinates of anomalous particles: (a) - type 1, $T$ = 1.35 K, $U_{top}$ = +100 V, $U_{bottom}$ = -110 V, cw laser power 42 mW; (b) - type 2, $T$ = 1.35 K, $U_{top}$ = -250 V, $U_{bottom}$ = +300 V, vertical arrow marks the moment when the cw laser power is changed from 45 to 235 mW.}
	\label{fig:XYvsTime}
\end{figure}

In addition to the collective oscillations induced by the surface waves, we observe also a small number of anomalous particles whose motion is strikingly different from that described above.
Two anomalous trajectories can be seen in Fig. \ref{fig:MapsTraj}(a) and one in Fig. \ref{fig:MapsTraj}(b).
We distinguish two types of anomalous particles.
The particles of type 1 oscillate with a larger amplitude and out of phase with the normal particles.
Their trajectory represents a sequence of circular loops 50--200 $\mu$m in diameter, as shown in Fig. \ref{fig:MapsTraj}(c).
The particles of the second type move more or less straight at approximately constant velocities, sometimes across the entire field of view of the camera.
The trajectory of such a particle typically consists of straight or slightly curved segments with sharp turning points, as can be seen in Figs. \ref{fig:MapsTraj}(a), (b) and (d).
In many cases this motion is quasiperiodic.
At the turning points, the particle abruptly changes the direction of motion.
In some cases, the velocities on different segments are different, but are well reproduced over many periods.
Both types of anomalous motion persist for the whole duration of the observation of the particular set of particles, sometimes up to several minutes.
$X(t)$ and $Y(t)$ curves typical for the type 1 and type 2 particles are shown in Figs. \ref{fig:XYvsTime} (a) and (b), respectively.

When an anomalous particle of type 2 moves through the cloud of trapped normal particles, it experiences numerous collisions.
The turning points are usually associated with some of those collisions.
In many other collisions, the anomalous particle keeps moving along the straight trajectory almost without any deviation.
Instead, the normal particles are pushed aside, make small loops and return to their equilibrium positions behind the passed anomalous particle.
Several such events can be seen along the straight track of the anomalous particle in Fig. \ref{fig:MapsTraj}(a).   

The power of the cw blue laser illuminating the trapped particles can influence the particle motion along the surface.
Higher laser power leads to a more active particle motion, lager amplitude of particle oscillations (see Fig. \ref{fig:XYvsTime}(b)) and a larger number of anomalous particles moving at higher velocities.

Particle motion in our experiment can be induced by surface waves and by a counterflow induced by the heating of the cell walls and particles themselves by the cw laser illumination.
However, both mechanisms result in a \textit{collective} motion of all particles and can not explain the individual anomalous motion of one or several particles within a large array.
It is expected that some number of quantized vortices is always present in superfluid He under the conditions of our experiments.
It has been demonstrated both theoretically \cite{KivotidesPRB2007,KivotidesPRB2008,ShuklaPRA2018} and experimentally \cite{BewleyN2006,GordonJETPL2007,FondaRSI2016} that solid nano- and microparticles become trapped by the vortices and move together with them.
We therefore suggest that the anomalous particles differ from the regular ones due to their binding to quantized vortices and their motion is driven by the dynamics of the vortex.

\begin{figure} [tbp]
	\includegraphics [width=\columnwidth] {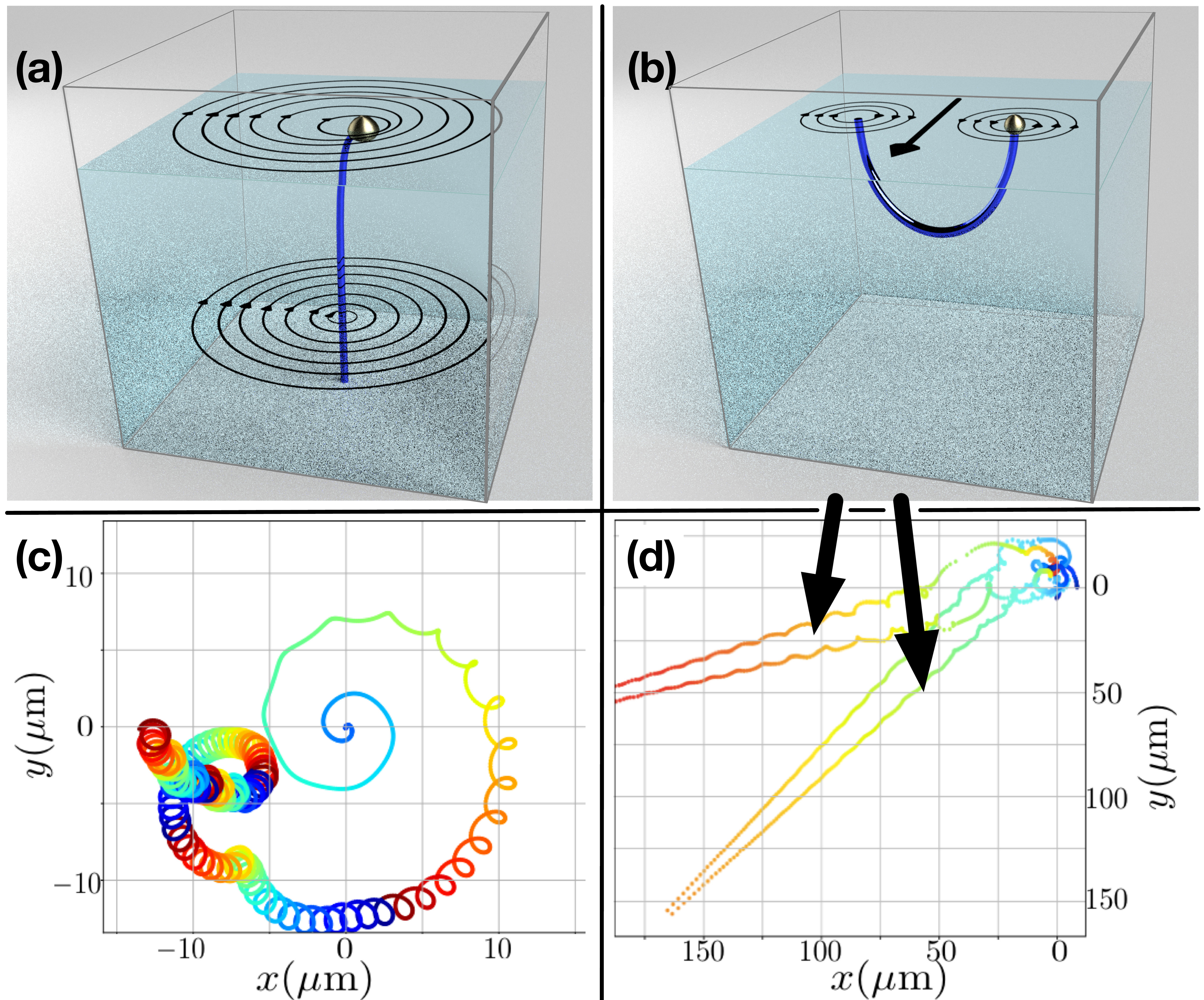}
	\caption{Schematic drawings of tracer particles trapped by quantized vortices: (a) linear vortex, (b) half-ring vortex. (c) Calculated particle trajectory corresponding to (a). (d) Calculated trajectories of two half-ring vortices created as a result of multiple reconnections of three linear vortices.}
	\label{fig:TrajCalc}
\end{figure}

Our theoretical model of the coupled dynamics of quantized vortices and particles is based on the vortex filament model \cite{MinedaPRB2013}.
The simplest configuration shown in Fig. \ref{fig:TrajCalc}(a) consists of a single vertical straight vortex filament and a particle attached together.
It is assumed that the upper end of the filament is connected at the center of the spherical particle and both move at the same speed.
The motion is excited by a collision between the straight filament at rest with a small vortex ring.

The trajectory of the particle at the surface following the excitation is shown in Fig. \ref{fig:TrajCalc}(c).
The collision leads to a transient dynamics, when the particle moves along a spiral trajectory away from the equilibrium position.
This spiral motion decays and is replaced by a smaller scale circular motion that does not depend on the details of the excitation and therefore is intrinsic for this system.
The mechanism can be summarized as follows.
The main part of the long vortex filament remains straight and vertical.
It creates a concentric velocity field $v_{s,BS}$ that drops inversely proportional to the distance from the center.
Small displacements of the particle together with the upper segment of the filament do not perturb this velocity profile significantly.
The total superfluid velocity $v_{s}$ is expanded into two terms: $v_{s} = v_{s,BS} + v_{s,LIA}$, where $v_{s,LIA}$ is obtained in a local induction approximation and is determined by a local curvature of the filament $s''$.
When the particle is kicked out off axis of the vortex velocity field, it shows the circular motion about a new equilibrium location which is determined by balancing the two forces acting on the vortex segment: the Magnus force $F_{M} \propto s' \times (\dot{s} - v_{s,BS})$ and the tension force $F_{t} \propto s''$. 
Here, $s$ is the filament and the prime and the dot symbols represent the derivatives with respect to arc length and time, respectively. 
The Magnus force acts outward and is dominant when $v_{s,BS}$ is large. 
On the other hand, the tension only depends on the local curvature of the filament.
As the particle gets farther the local curvature of the vortex becomes larger and the tension force that pulls the particle backward becomes dominant.
The circular motion strongly resembles the experimentally observed anomalous dynamics of type 1 (see Fig. \ref{fig:MapsTraj}(c)), although the computed trajectory radii are significantly smaller than experimental. 
As discussed in \cite{InuiJLTP2018}, the radius increases with the size of the particle.
In the experiment, the particle heated by the laser light may become surrounded by a shell of normal fluid He which may lead to the increased effective particle size and a larger trajectory radius.

In order to explain the anomalous particle motion of type 2, we have to consider more complex systems including several quantized vortices.
In particular, a pair of linear vortices orthogonal to the free surface and rotating in two opposite directions is expected to move parallel to the surface at a constant speed that is inversely proportional to the distance between the vortices $d$: $v_{d} = \Gamma /2\pi d$.
$\Gamma$ here is the quantum of the circulation: $\Gamma = h/M_{He} = 9.97 \times 10^{-4}$ cm$^{2}$/s.
Such complexes, also known as quantized vortex dipoles, have been predicted and investigated theoretically \cite{CrasovanPRA2003,LiPRA2008,GeurtsPRA2008}.
More recently, these interesting macroscopic quantum objects were observed in atomic Bose-Einstein condensates (BEC) \cite{FreilichS2010,NeelyPRL2010,MiddelkampPRA2011,KwonPRA2015} and in exciton-polariton condensates \cite{RoumposNP2010,NardinNP2011}.
Vortex dipoles in superfluid He have been predicted theoretically  \cite{RoraiPRB2013} and the interaction and trapping of particles by vortex dipoles and larger clusters of quantized vortices was modeled numerically in \cite{ShuklaPRA2018}. 
However, no experimental observations have been reported up to date.

Our calculations demonstrate that two counter-rotating linear vortices at a close distance become unstable.
The two filaments approach each other and reconnect in one or several places producing several vortex rings of different sizes.
Remarkably, the upper segments of the two filaments touching the free surface upon the reconnection transform into a surface-bound vortex of a new type: a half-ring with a radius $R$ equal to one half of the distance between the two parent vortices and with both ends terminated at a free surface.
Similar to the motion of vortex dipoles and vortex rings in the bulk \cite{DonnellyBook}, the half-ring vortex moves along the direction of the ring axis, with a velocity $v_{hr} = \Gamma /4\pi R$.
With the both ends terminated at the free surface, the half-ring maintains a vertical orientation and moves parallel to the surface.
This translational motion is accompanied by the oscillations of the half-ring radius, that depend on the details of the collision and reconnection of the two initial vortices.
Fig. \ref{fig:TrajCalc}(d) shows the motion of two half-ring vortices created as a result of the reconnections of three linear vortices.
The trajectory of each half-ring is represented by the traces of its two tips along the surface.

The scenario outlined above allows us to assign the anomalous particle motion of type 2 to the particles trapped by half-ring vortices traveling along the free surface of superfluid He.
The particle of a sub-micron diameter is supposed to be trapped at one end of the vortex, as shown in Fig. \ref{fig:TrajCalc}(b).
Experimentally measured particle velocities fall in the range of 0.1--10 cm/s, which corresponds to the radii of the half-ring vortices of 0.1--10 $\mu$m.
The nearly constant velocity of many type 2 particles implies that the radius of the half-ring in each case remains constant over a long period of time.
When the particle attached to one end of the half-ring hits an obstacle such as another charged particle or the inhomogeneity of the external electric field at the edge of the capacitor, it slows down or stops abruptly and the other end of the half-ring moves around it.
As a result, the half-ring changes its orientation and starts moving away from the obstacle.
Similar effect can be produced by collisions of the complex with other co-rotating vortices.
The collisions may also lead to the abrupt change of the half-ring radius and the corresponding velocity modulus.
The particle-vortex interaction is sufficiently strong that the particle driven by the vortex is able to overcome the Coulomb repulsion and approach close to other particles and even kick them out from their equilibrium positions in the trap.

Our observations clearly demonstrate that upon an increase of the illuminating light intensity some regular particles start moving as anomalous of type 2.
We suggest that the heating of individual particles by the intense laser light induces a strongly localized radial counterflow around the particle which exceeds the critical velocity and leads to the creation of new vortices pinned at the particle.
For some particles, at a high laser power we observe frequent and chaotic changes of the direction and magnitude of their velocity which may be a signature of multiple vortices interacting with the particle.

Even more intriguing is the behavior of the particles that upon the increase of the light intensity increase their speed, but retain their highly regular periodic character of motion, as shown in Fig. \ref{fig:XYvsTime} (b).
The heating of the particle thus results in the shrinking of the half-ring vortex bound to it.
The effect can be attributed to the heat-induced counterflow around the particle, in which the superfluid component is moving towards the particle and the normal component - away from it.
The free end of the superfluid half-ring vortex is thus pushed towards the particle. 
It is thus possible to manipulate the size of the half-ring vortices and control their motion by shining the laser light onto the trapped particle, which opens new perspectives for experiments.

In summary, we have developed a new method for visualizing the motion of quantized vortices terminating at a free surface of superfluid He that relies on the usage of electrically charged tracer particles trapped at the surface.
We interpret our observations in terms of two types of surface-bound vortices with the characteristic dynamics: a linear vertical filament and a half-ring. 
The former remains pinned at the bottom, with the upper end and the attached particle moving along a circular trajectory $\approx$100 $\mu$m in diameter.
The latter moves together with the trapped particle along a straight line and turns around abruptly when meets an obstacle.
The vortices interact with the laser light via the heating of the trapped particle and the resulting local counterflow.

\begin{acknowledgments}
	
This work was supported by JSPS KAKENHI grants No JP24000007, JP17H01145, and JP17K05548.
	
\end{acknowledgments}


\end{document}